\documentclass[aps, prl, twocolumn,showpacs,preprintnumbers,amsmath,amssymb]{revtex4}

\usepackage{graphicx}
\usepackage{dcolumn}
\usepackage{bm}

\begin{document}

\title{Perturbations can enhance qauntum search}

\author{Joonwoo Bae$^{1}$} 
\email{Joonwoo.Bae@upc.es}
\author{ Younghun Kwon$^{1,3}$}
\email{yyhkwon@hanyang.ac.kr}
\affiliation{$^1$ Department of Physics, Hanyang University, \\Ansan, Kyunggi-Do, 425-791, South Korea\\}

\affiliation{$^2$ Department of Physics, Unversity of Rochester, Bausch and Lomb Hall, P.O. Box 270171,600 Wilson Boulevard, Rochester, NY 14627-0171 \\}

\date{\today}

\begin{abstract}
In general, a quantum algorithm wants to avoid decoherence or perturbation, since such factors may cause errors in the algorithm. In this letter, we will supply the answer to the interesting question: can the factors seemingly harmful to a quantum algorithm(for example, perturbations) enhance the algorithm? We show that some perturbations to the generalized quantum search Hamiltonian can reduce the running time and enhance the success probability. We also provide the narrow bound to the perturbation which can be beneficial to quantum search. In addition,  we show that the error induced by a perturbation on the Farhi and Gutmann Hamiltonian can be corrected by another perturbation. 
\end{abstract}

\pacs{03.67.Lx}
\keywords{Quantum Search, Hamiltonian}

\maketitle

Quantum computation and information theory has been in the spotlight, with the expectation that a quantum computer may possess a surprising computational power and a quantum information processor may guarantee the security ,which could be broken in a known classical protocol. The Shor's quantum factorization algorithm of the exponential speedup and the Grover's quantum search algorithm of the quadratic speedup were examples which prove quantum computational power.\cite{1}\cite{2} Furthermore, the recent proposal for quantum search based on Hamiltonian evolution showed the $O(1)$ running time.\cite{6} The no-cloning theorem also implied that a quantum protocol is superior to a classical one in security. \\
 The characters of a quantum machine may be concluded as the three factors; quantum superposition, quantum interference, and quantum entanglement.\cite{6}\cite{10} Moreover, the processing procedure of the machine should be unitary. These are the ingredients to discriminate a quantum algorithm from a classical one. However, there are difficulties in implementation of a quantum computer. For example, some decoherence or some perturbation on a quantum algorithm can induce a fatal error.\cite{4} 
 Then, the following interesting question arises: Is there a case that the factors considered harmful to a quantum algorithm may enhance the algorithm? As we will explain it, the answer is  "yes". The discovery that non-ideal ingredients, e.g., decoherence, perturbation, noise, or error, which have been considered to be negative to a quantum algorithm, improve a given quantum algorithm is indeed a good news. The unexpected result may provide a novel point of view in developing the ways to implement a quantum algorithm. 
 In this letter, we consider perturbations on a quantum search Hamiltonian. We then show some perturbations to  the Hamiltonian can enhance quantum search. The quantum search algorithm based on Hamiltonian evolution was proposed by Farhi and Gutmann.\cite{3} They showed that the following Hamiltonian

\begin{eqnarray}
 H=E(|w \rangle \langle w| + |\psi \rangle \langle \psi|)  \nonumber \\
\end{eqnarray}

finds the target state $|w \rangle$ of the state $|\psi \rangle$ superposed with $N$ states. After $O(\sqrt{N})$ evolution times, the target state is obtained with probability one. Oshima showed that if the Hamiltonian is initialized as follows

\begin{eqnarray}
 H=E|w \rangle \langle w| + E^{'}|\psi \rangle \langle \psi|  \nonumber \\
\end{eqnarray}

with $E \neq E^{'}$, then the running time becomes $O(N)$.\cite{7} This implies that inaccurately initialized energies make quantum search fail.  Thus perturbations on the energy of the Hamiltonian can be fatal to the algorithm. \\
 The generalized quantum search Hamiltonian was recently presented, which is\cite{5}

\begin{eqnarray}
 H=E(|w \rangle \langle w| + |\psi \rangle \langle \psi|)  + \epsilon (e^{i\phi}|w \rangle \langle \psi| + e^{-i \phi} |\psi \rangle \langle w|) \nonumber \\
\end{eqnarray}

where $E$ and $\epsilon$ are constants in unit of energy with $E\geq \epsilon$, and $\phi$ is a constant phase. Using the initial state as $|\psi(t=0) \rangle = x |w \rangle + \sqrt{1-x^{2}}|r \rangle$, where $x = \langle w|\psi \rangle  \approx 1/\sqrt{N}$, we have the success probability at the proper-time $T_{0}$ 

\begin{eqnarray}
P_{0}(T_{0}) & = &|\langle w |e^{-iHt}|\psi(t=0) \rangle|^{2} \nonumber \\
& = & (1-x^{2}) + \frac{x^{2}(Ex+\epsilon cos \phi)^{2}}{(Ex+\epsilon cos\phi)^{2}+(1-x^{2})\epsilon^{2}sin^{2}\phi} \nonumber \\
& \geq & 1-x^{2} \approx 1-\frac{1}{N}\nonumber \\
\end{eqnarray}

 Thus, the lower bound of the success probability is $1-O(1/N)$. The running time of the algorithm is as follows:

\begin{eqnarray}
T_{0} = \frac{\pi}{2}\frac{1}{[(Ex+\epsilon cos\phi)^{2}+(1-x^{2})\epsilon^{2}sin^{2}\phi]^{1/2}}\nonumber \\
\end{eqnarray}

This Hamiltonian finds the target state at least in $O(\sqrt{N})$ times and with probability $1-(1-\delta_{\phi, n\pi})O(1/N)$, $n \in \mathbb{Z}$.  Now let us consider perturbations on the Hamiltonian under the assumption that the evolution $e^{-iHt}$ is unitary.($\hbar=1$ throughout) The perturbations to be considered are on the phase $\phi$ and the energy $E$. Then we will show that, surprisingly, some perturbations can reduce the running time and boost the probability of finding the target.  \\
  If  we consider a perturbation on the phase $\phi$. it can be induced as follows :

\begin{eqnarray}
 H=H_{0}+H_{1} \nonumber \\
\end{eqnarray}

where

\begin{eqnarray}
H_{0} &=& E(|w \rangle \langle w| + |\psi \rangle \langle \psi|) \nonumber \\ && + \epsilon_{0} (e^{i\phi_{0}}|w \rangle \langle \psi| + e^{-i \phi_{0}} |\psi \rangle \langle w|) \nonumber \\
H_{1} &=& \epsilon_{1} (e^{i\phi_{1}}|w \rangle \langle \psi| + e^{-i \phi_{1}} |\psi \rangle \langle w|) \nonumber \\
\end{eqnarray}

The Hamiltonian $H_{0}$ is the original quantum search Hamiltonian. The Hamiltonian $H_{1}$ is the Hamiltonian creating the perturbation on the phase $\phi_{0}$, as follows:

\begin{displaymath}
H=E( |w \rangle \langle w| + |w \rangle \langle s|) + \epsilon^{'}(e^{i\varphi} |w \rangle \langle \psi| + e^{-i\varphi}|\psi \rangle \langle w|)\nonumber \\
\end{displaymath}

where

\begin{eqnarray}
 \epsilon^{'} &=& \sqrt{\epsilon_{0}^{2}+\epsilon_{1}^{2}+2\epsilon_{0}\epsilon_{1}cos(\phi_{0}-\phi_{1})} \nonumber \\ 
 \varphi &=& cos^{-1}(\frac{\epsilon_{0}cos\phi_{0}+\epsilon_{1}cos\phi_{1}}{\epsilon^{'}}) \nonumber \\
\end{eqnarray}

In this case, the running time may be reduced if the energy $\epsilon_{0} \leq \epsilon^{'}$. The success probability may not be improved, but the infimum is still $1-O(1/N)$. The probability-boost depends only on the phase $\varphi$. Thus, we have learned that the perturbation on the phase may provide a reduced running time and an improved success probability.\\
 A perturbation on the energy $E$ of the Hamiltonian is quite different from that on the phase we considered. Contrary to the case that a perturbation on the phase does not corrupt the quantum search, a perturbation on the energy can spoil it. Since we have assumed that evolution of the Hamiltonian is unitary,  it is sufficient to consider  the perturbation on the term $E(|w \rangle \langle w|+ |\psi \rangle \langle \psi|)$. Thus the perturbed Hamiltonian is 

\begin{eqnarray}
 H=H_{0}+H_{2} \nonumber \\
\end{eqnarray}

where

\begin{eqnarray}
H_{0} &=& E(|w \rangle \langle w| + |\psi \rangle \langle \psi|) \nonumber \\ && + \epsilon (e^{i\phi}|w \rangle \langle \psi| + e^{-i \phi} |\psi \rangle \langle w|) \nonumber \\
H_{2} & = & 2\triangle |w \rangle \langle w|\nonumber \\
\end{eqnarray}

Then, at the proper-time $T_{e}$, the probability to find target is

\begin{eqnarray}
P_{e}(T_{e}) &=& | \langle w |e^{-iHt}|\psi \rangle |^{2}\nonumber \\
& = &(1-x^{2})+ \frac{A_{e}}{M_{e}^{2}}\nonumber \\
\end{eqnarray}

where

\begin{eqnarray}
A_{e} &=& -(1-2x^{2})\triangle^{2}+2x^{3}(Ex+\epsilon cos\phi)\triangle  \nonumber \\&&+x^{2}(Ex+\epsilon cos\phi)^{2} \nonumber \\
M_{e} &=& [((Ex+\epsilon cos\phi)^{2}+ \epsilon^{2}sin^{2}\phi)(1-x^{2})\nonumber \\&&+((Ex+\epsilon cos\phi)x+\triangle)^{2} ]^{1/2} \nonumber \\
\end{eqnarray}

The running time is 

\begin{eqnarray}
T_{e} = \frac{\pi}{2} \frac{1}{M_{e}} \nonumber \\
\end{eqnarray}

We here provide the exact relation between the running time and the probability.(Here subscript e and 0  mean the case of perturbation and  that of non-perturbation respectively)

\begin{eqnarray}
\frac{\pi^{2}}{4}(P_{e} - P_{0}) = A_{e}T_{e}^{2} - A_{0}T_{0}^{2} \nonumber \\
\end{eqnarray}

where

\begin{eqnarray}
A_{0} & = & x^{2}(Ex+\epsilon cos\phi)^{2}\nonumber \\
\end{eqnarray}

 The arising interest is whether the success probability and the running time can be simultaneously improved by the perturbation $\triangle$ or not. The following shows that the running time can be reduced.\\

$Remark1.$ In the case of $Ex + \epsilon \cos \phi \geq 0$, we have $T_{e} \leq T_{0}$ if $\triangle \geq 0$ or $\triangle \leq -2x(Ex+\epsilon \cos \phi)$. In the case of $Ex+\epsilon \cos \phi \leq 0$, we have $T_{e}\leq T_{0}$ if $\triangle \leq 0$ or $\triangle \geq -2x(Ex+\epsilon \cos \phi)$. \\

The success probability can be also improved as follows. \\

$Remark2.$ For the case of $Ex + \epsilon \cos \phi \geq 0$, we have $P_{e} \geq P_{0}$ if $\triangle \in [0,\beta_{>}]$. For the case of $Ex+\epsilon \cos \phi \leq 0$, we have $P_{e} \geq P_{0}$ if $\triangle \in [\beta_{<},0]$. \\

\begin{eqnarray}
\beta & = & \frac{2x^{3}(Ex+\epsilon cos \phi)\epsilon^{2}sin^{2}\phi}{(Ex+\epsilon cos\phi)^{2}+(1-2x^{2})\epsilon^{2}sin^{2}\phi} \nonumber \\
\end{eqnarray}

$\beta_{>}$ and $\beta_{<}$ denote a positive and negative value of  $\beta$ respectively. The remarks imply that 
we can search a target state with a slightly improved speedup and a boosted probability, by perturbing the energy as the amount bounded by $[\beta_{<}, \beta_{>}]$. We, however, note that the probability-boost does not occur under some situations. If the energy $\epsilon$ is zero, or if the phase is given as $\phi = cos^{-1}(-Ex/\epsilon)$ or $\phi = n\pi$, then we cannot achieve an improved probability. In the case of $\epsilon =0 $, the probability becomes one in the proper time, so there cannot be a probability-boost. The phase $\phi = cos^{-1}(-Ex/\epsilon)$ and $\phi = n\pi$ make $\beta$ zero, so there is no gain by the perturbation. Also, we note that the running time cannot be reduced if $\phi = cos^{-1}(-Ex/\epsilon)$. \\
 We here observe that the perturbation on the energy is classified by the factor $Ex+\epsilon \cos \phi$. This implies that a beneficial perturbation depends on the phase $\phi$. That is, the perturbations on the energy are  closely related with that on the phase. \\

$Classification ~~ of ~~ Perturbation$\\

1. $Ex+\epsilon \cos \phi \geq 0$\\
i) If $\triangle \in [0,\beta_{>}]$, then we have $P_{e}\geq P_{0}$ and $T_{e}\leq T_{0}$. Therefore, both are improved.\\
ii) If $\triangle \geq \beta_{>}$ or $\triangle \leq -2x(Ex+\epsilon \cos \phi)$, then we have $P_{e}\leq P_{0}$ and $T_{e}\leq T_{0}$. Therefore, the running time is improved only.\\
iii) If $\triangle \in [-2x(Ex+\epsilon \cos\phi),0]$, then both are corrupted.\\

2. $Ex+\epsilon \cos \phi \leq 0$\\
i) If $\triangle \in [\beta_{<},0]$, then we have $P_{e}\geq P_{0}$ and $T_{e}\leq T_{0}$. Therefore, both are improved.\\
ii) If $\triangle \leq \beta_{<}$ or $\triangle \geq -2x(Ex+\epsilon \cos \phi)$, then we have $P_{e}\leq P_{0}$ and $T_{e}\leq T_{0}$. Therefore, the running time is improved only.\\
iii) If $\triangle \in [0, -2x(Ex+\epsilon \cos\phi)]$, then both are corrupted.\\

Thus 
we have shown that a perturbed energy and a perturbed phase can enhance quantum search, and also provided the bound conditions for good and bad perturbations. It is quite remarkable that there is a bound where the probability and the running time are both improved by the perturbation $\triangle$. We here note that the bounds is as narrow as an amount of $O(1/N)$.  \\
 As we stated, the perturbed Farhi and Gutmann Hamiltonian, $(E+\triangle)|w\rangle \langle w|+E|\psi \rangle \langle \psi|$, fails quantum search since the evolution time becomes $O(N)$. We now know that, if the Farhi and Gutmann Hamiltonian is perturbed, then the error can be corrected by the perturbation $\epsilon(e^{i\phi}|w\rangle\langle \psi|+e^{-i\phi}|\psi \rangle \langle w|)$. In other words, the Farhi and Gutmann Hamiltonian can be improved by attaching the additional term.\\

\section*{Acknowledgement}
J. Bae is supported in part by the Hanyang University Fellowship and Y. Kwon is supported in part by the Fund of Hanyang University.

\end{document}